\documentclass{iau}

\usepackage{amsmath}
\usepackage{graphicx}
\usepackage{multirow}
\usepackage{wrapfig}

\newcommand{\2}{$_2$}
\newcommand{\3}{$_3$}
\newcommand{\MN}{CH\3OH}
\newcommand{\MF}{HCOOCH\3}
\newcommand{\TFA}{H\2CS}
\newcommand{\cycCH}{c-C\3H\2}
\newcommand{\MT}{CH$_4$}
\newcommand{\CO}{C$^{17}$O}

\newcommand{\iras}{IRAS 16293$-$2422}
\newcommand{\irass}{IRAS 15398$-$3359}

\newcommand{\apj}{{\it ApJ}}
\newcommand{\apjl}{{\it ApJL}}
\newcommand{\apjs}{{\it ApJS}}
\newcommand{\aap}{{\it A\&A}}
\newcommand{\mnras}{{\it MNRAS}}
\newcommand{\pasp}{{\it PASP}}

\begin{document}

\lefttitle{Yoko Oya}
\righttitle{Chemical Diagnostics for Tracing the Physical Structures in Disk-Forming Regions}

\jnlPage{1}{7}
\jnlDoiYr{2021}
\doival{10.1017/xxxxx}

\aopheadtitle{Proceedings IAU Symposium}
\editors{C. Sterken,  J. Hearnshaw \&  D. Valls-Gabaud, eds.}

\title{Chemical Diagnostics for Tracing the Physical Structures in Disk-Forming Regions of Young Low-Mass Protostellar Sources}

\author{Yoko Oya}
\affiliation{Yukawa Institute for Theoretical Physics, Kyoto University}

\begin{abstract}
To understand the chemical origin of the Solar system, the chemical evolution along the star/planet formation is a key issue. 
Extensive observational studies have demonstrated a chemical diversity in young low-mass protostellar sources so far. 
Furthermore, chemical differentiations in the vicinity of the protostars have recently been reported. 
This suggests that molecular distribution is sensitive to a change in the physical conditions associated with disk formation. 
Some kinds of molecular lines, 
especially Sulfur-bearing species, 
are therefore prospected to work as molecular markers to highlight particular structures of disk-forming regions. 
 
Conversely, detailed physical characterization is essential for elucidating the chemical evolution occurring there. 
Machine learnings may help us to disentangle the observed structures. 
Angular momentum of the gas is the key topic to understand the structure formation, which is also essential to the integration of the chemical and physical characterization.
\end{abstract}
%It has recently been recognized that some molecular lines tend to trace the mid-plane of a disk/envelope system and others trace its surface. 
%The traced kinematic structure may differ among these molecular emissions, 
%and thus, their careful analyses are essential for further observational studies in disk-forming regions. 
%
%Different kinematic structures are separately traced by different molecular tracers, 
%and thus, their careful analyses are essential for further studies in disk-forming regions. 


\maketitle

\section{Introduction} \label{sec:intro}

To explore the chemical origin of the Solar system, the chemical evolution along the star/planet formation has been extensively examined. 
For instance, the chemical compositions in its earlier phase (protostellar cores) have been observed at the radio wave-length. 
That in the later phase (planetary systems) has been examined by sample return missions within the Solar system. 
The intermediate phase is that for the formation of protostellar/protoplanetary disks, 
where a surprisingly wide variety of organic molecules are produced. 
While theoretical studies have been performed extensively for this phase, 
the observational studies had a longstanding missing link. 
Atacama Large Millimeter/submillimeter Array (ALMA) has given large impacts on this field in this decade. 

In spite of these efforts, it is still observationally controversial 
{\it `when and how disks are formed around a newly-born protostar'} 
and {\it `what kinds of molecules are delivered into disks'}. 
These important issues in astrophysics and astrochemistry are also related with the diversity in planetary systems.

\section{Organic Molecules in Disk-Forming Regions} \label{sec:chem}

%\subsection{Chemical Diversity among Sources} 

It is known that low-mass protostellar cores show a chemical diversity \citep[e.g.][]{Lefloch2018}.  
For instance, two distinct cases are the hot corino chemistry and warm carbon-chain chemistry (WCCC). 
Hot corinos are rich in complex organic molecules 
(COMs; e.g. \MN, CH$_3$CHO) \citep[e.g.][]{Schoier2002, Ceccarelli2004}. 
WCCC sources are rich in unsaturated hydrocarbons (e.g. CCH, \cycCH) \citep[e.g.][]{Sakai2013}. 
{\it How these diversities are evolved toward planetary systems} is an important clue to the material origins of the Solar system.

\subsection{Chemical Differentiation within One Source} \label{sec:chem-diff}

The chemical diversity in protostellar cores was recently found to be interpreted as a single picture in disk-forming regions 
\citep[][]{OyaSpringer}. 
The above two distinct cases have been thought to be exclusive to each other; 
however, some protostellar sources were found to show both of these chemical characteristics. 
%Moreover, it is interesting that the two chemical characteristics are spatially resolved. 

\begin{figure}
\begin{center}
\includegraphics[bb = 0 0 550 500, scale = 0.65]{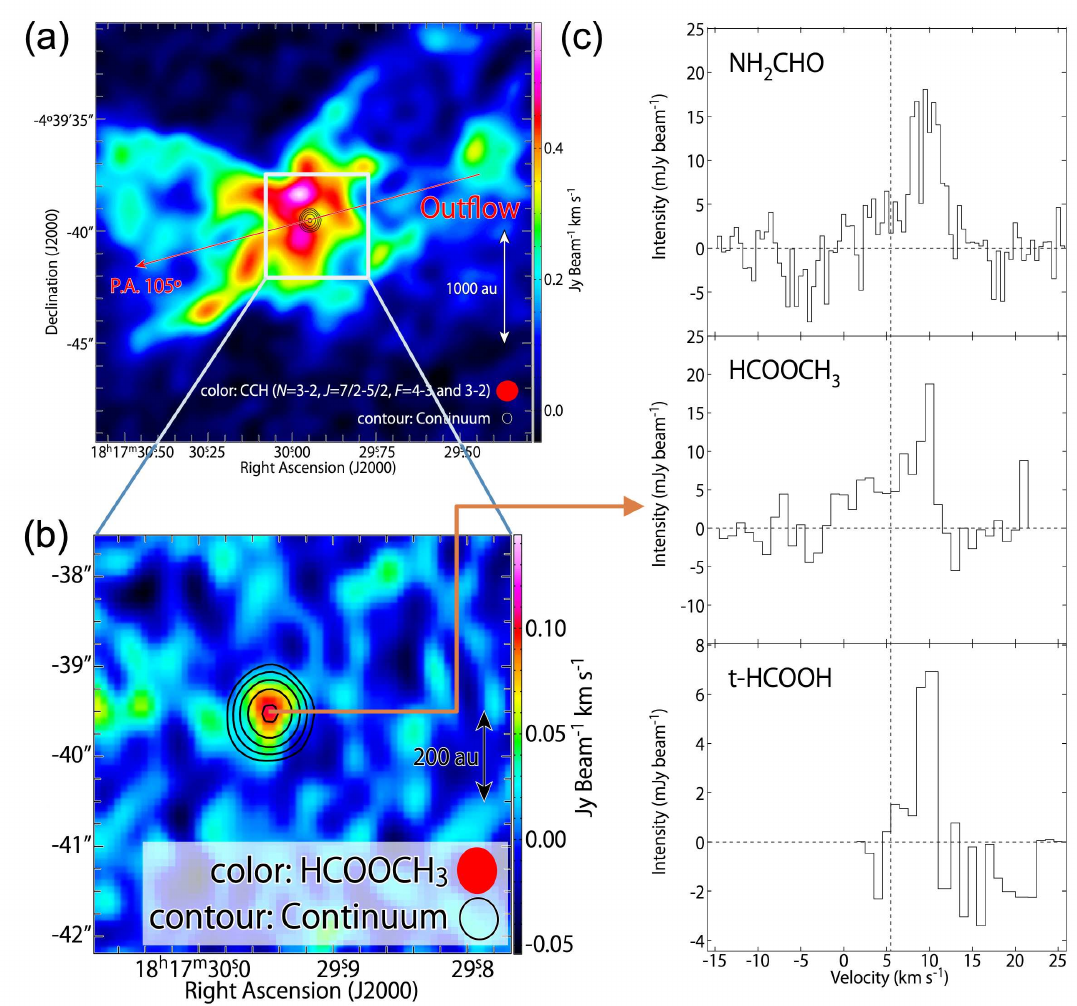}
\caption{%
(a, b) 
Integrated intensity maps of the CCH emission (a; color) 
and the HCOOCH$_3$ emission (b; color) 
toward L483 observed with ALMA. 
Contours represent the 1.2 mm continuum emission. 
Distributions of the emission of hydrocarbons are spread over $10^3$ au, 
while those of COMs are concentrated to the protostellar position. 
(c) Spectra toward the protostellar position showing the detection of COMs. 
Panels are taken from \citet{Oya2017} with modifications.
}
\end{center}
\label{fig:L483}
\end{figure}

L483 is an example with such a `hybrid' chemical characteristics (Figure \ref{fig:L483}). 
L483 is a dark cloud core located in Aquila Rift \citep[$d=200$ pc;][]{Rice2006}, 
which harbors a Class 0 low-mass protostellar source IRAS 18148$-$0440 \citep{Fuller1995}. 
ALMA observations revealed that 
the emissions of hydrocarbons are detected in the extended envelope gas and the outflow lobes in this source \citep{Oya2017, Oya2018-L483}. 
Meanwhile, the emissions of COMs were detected only in the vicinity of the protostar. 
It is interesting that the two different chemical characteristics are spatially separated.

%\begin{wrapfigure}[28]{o}[0pt]{0.55\textwidth}
%\end{wrapfigure}

\begin{figure}
\begin{center}
\includegraphics[bb = 0 0 650 250, scale = 0.52]{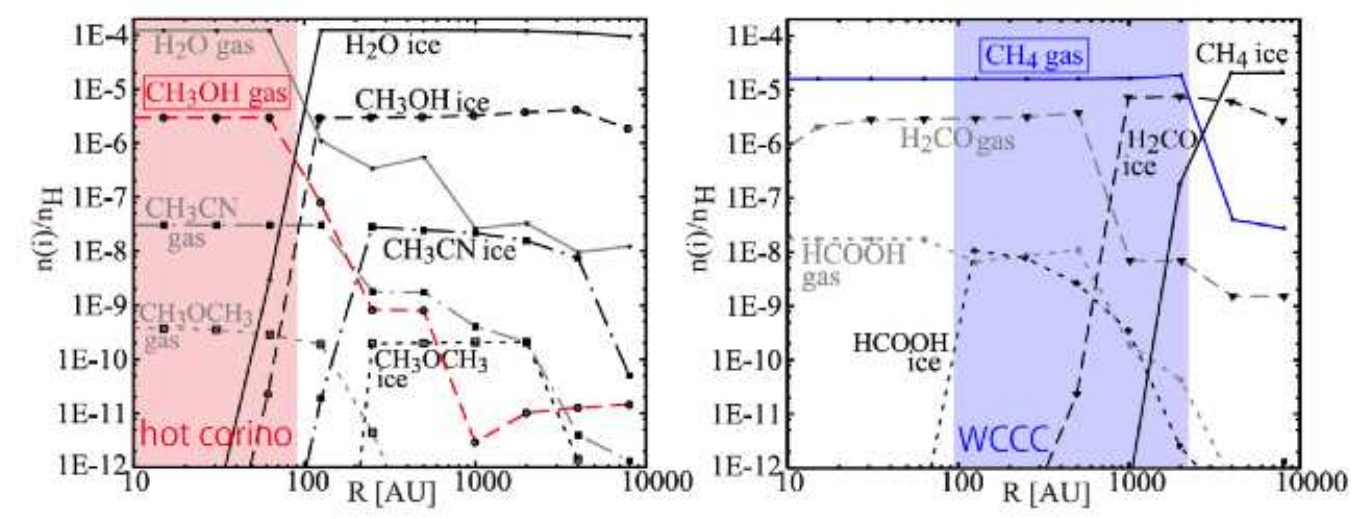}
\caption{%
Plots of the molecular abundances as a function of the distance to the protostar. 
Plots are taken from \citet{Aikawa2008} with modifications. 
}
\end{center}
\label{fig:chemmodel}
\end{figure}

The chemical differentiation of organic molecules found in L483 
has been confirmed in other sources with the hybrid chemical characteristics \citep[B335, CB68;][]{Imai2019, Imai2022}. 
These observational results are actually consistent with a chemical model for a dynamically collapsing core reported by \citet{Aikawa2008}. 
Figure \ref{fig:chemmodel} shows the radial profiles of the molecular abundances in the chemical model. 
\MN\ is seen in the gas-phase at a distance $<10^2$ au from the protostar; 
it is sublimated from dust grains in the hot regions near the protostar, as known as the hot corino activity. 
Meanwhile, \MT\ is seen in the gas-phase in the warm region 
with a distance from $10^2$ au to $10^3$ au from the protostar. 
The difference in these molecular distributions in the gas-phase can be attributed to the difference in their desorption temperatures; 
\MN\ and other COMs are thought to be sublimated from the dust grains with water at a dust temperature $>100$ K, 
while \MT\ at a dust temperature $>25$ K. 

\begin{figure}[t]
\begin{center}
\includegraphics[bb = 0 0 700 500, scale = 0.5]{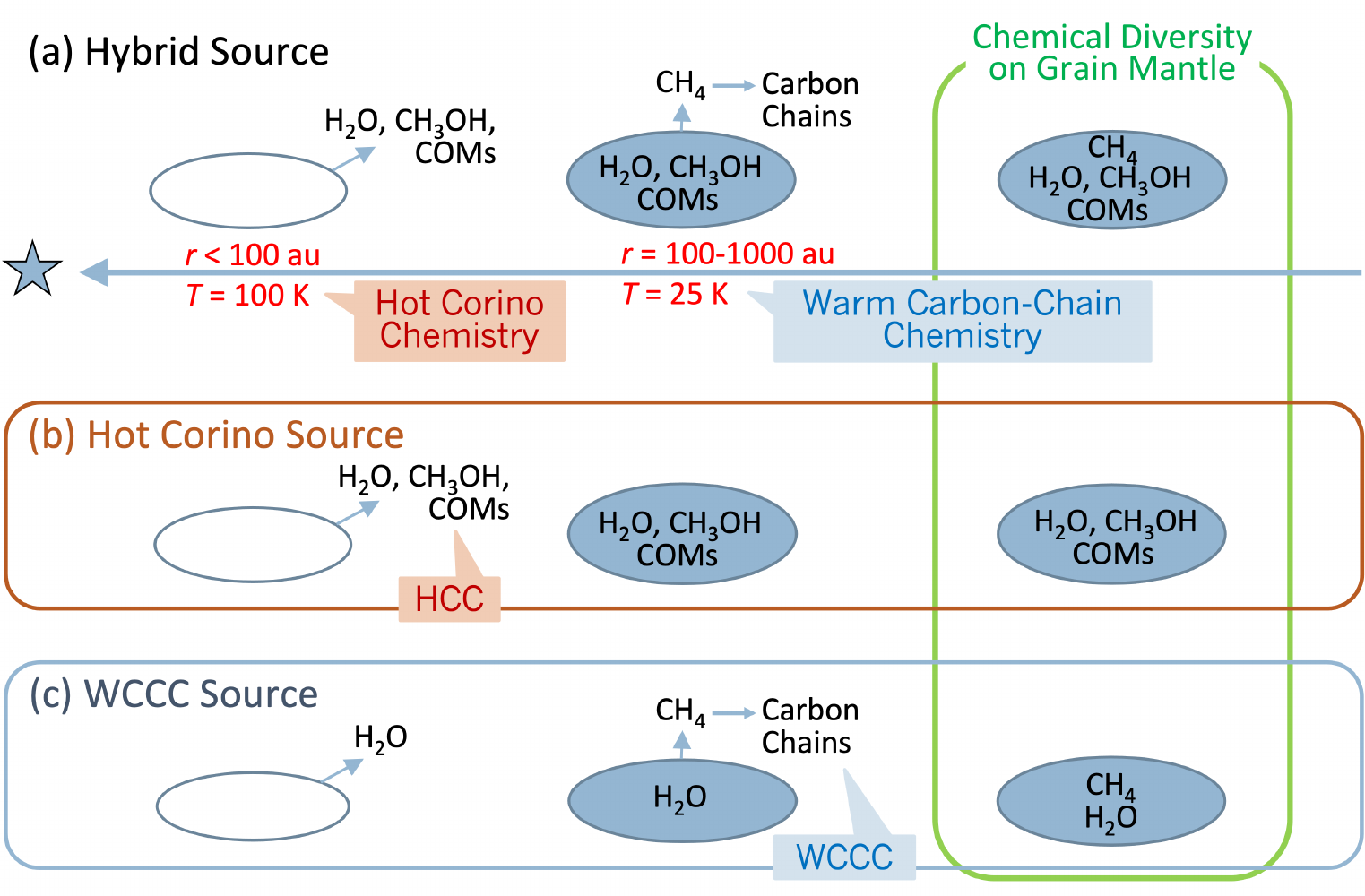}
\caption{%
Schematic illustration of the sublimation of molecules from dust grains. 
}
\end{center}
\label{fig:sublimation}
\end{figure}

Figure \ref{fig:sublimation} shows a simplified view for the sublimation of organic molecules from dust grains. 
The case for sources with the hybrid chemical characteristics is shown in Figure \ref{fig:sublimation}(a). 
First, various molecular species exist in grain mantles, as shown at the right end in this panel. 
Grains fall into a warm region ($T>25$ K), leading the sublimation of \MT.  
Enhancement of \MT\ in the gas-phase triggers efficient production of hydrocarbons (i.e. WCCC). 
Then, grains fall into a hot region ($T>100$ K), and water and COMs are sublimated (i.e. hot corino activity). 
This can be regarded as the basic picture.

\subsection{Chemical Variety among Sources} \label{sec:chem-var}

The situation described above depends on the initial chemical condition of dust grains. 
Figure \ref{fig:sublimation}(b) shows a case where \MT\ is deficient in dust grains. 
The abundance of \MT\ in the gas-phase is depressed in a warm region, 
while COMs are sublimated in a hot region; 
this case corresponds to hot corino sources. 
Alternatively, a case deficient in COMs is shown in Figure~\ref{fig:sublimation}(c). 
Hydrocarbons can be produced from \MT\ sublimated in a warm region, 
while the enhancement of abundance of COMs is not expected to be significant in a hot region; 
this case is regarded as WCCC sources. 
In summary, the chemical diversity in protostellar sources are likely originated from that on a grain mantle, 
which is expected to be determined according to the physical and chemical conditions in the prestellar stage. 

The chemical composition of a grain mantle mainly depends on 
what kinds of molecules deplete on a grain 
and how they are accumulated on a grain during the cold starless phase. 
More specifically, 
it is proposed to depend on the duration time of the starless-core phase. %\citep[][]{Sakai2013}. 
%\citet{OyaSpringer} suggested the standard case 
%where carbon atoms and CO molecules are adsorbed onto dust grains during the core collapse. 
%Their composition on dust grains is expected to suffer from 
%what rate of carbon atoms are converted to CO molecules before their adsorption. 
If the duration time is long enough, 
a significant rate of carbon atoms are converted to CO molecules in the gas phase before their adsorption, 
and produced CO molecules are accumulated on dust grains. 
Then, CO can form a variety of COMs via dust surface reactions, 
which will show the hot corino chemistry in the vicinity of a protostar. 
On the contrary, 
carbon atoms are depleted onto dust grains as they are, 
if the duration time is too short to convert them to CO molecules. 
Then, carbon atoms will be hydrogenated to \MT\ on dust surface, 
which will trigger WCCC. 
The duration time can be moderate; 
both of carbon atoms and CO molecules will exist on dust grains, 
resulting in the hybrid chemistry (Figure~\ref{fig:sublimation}a). 
The peculiar chemical characteristics of hot corino sources and WCCC sources have extensively been studied so far. 
However, their intermediate case, i.e. the hybrid chemistry, is suggested to be the standard case \citep{Oya2017}. 

The above picture is regarded as the competition between the time scales for the chemical equilibrium and the adsorption of molecules 
after the UV shielding \citep[][]{Higuchi2018, OyaSpringer}; 
the former is roughly estimated to be $\sim3 \times 10^5$ yr for a wide range of the H\2\ density, 
while the latter is inversely proportional to the H\2\ density. 
The balance of these time scales is expected to suffer from environmental effects. 
For instance, 
a strong UV radiation resets the chemical composition, 
and then carbon atoms are expected to tend to be adsorbed onto dust grains before they are converted to CO molecules. 
In a cloud supported by strong turbulence, 
the chemical equilibrium is expected to be achieved in the gas phase during the relatively slow cloud collapse. 
This qualitative understanding is actually consistent 
with the observational study for the 36 Class 0/I protostellar sources in the Perseus by \citet{Higuchi2018}.

\section{Chemistry Associated with Physical Structures} \label{sec:phys}

In practice, the chemical differentiation described in Section~\ref{sec:chem-diff} is much more entangled with the actual physical structures. 
An interesting case was reported for \iras\ in Ophiuchus \citep[$d\sim140$ pc][]{OrtizLeon2017}. 
\iras\ is a Class 0 low-mass protostellar binary source, consisting of Source A and Source B. 
Both of them are famous as hot corinos rich in COMs \citep[e.g.][]{vanDishoeck1995, Cazaux2003}. 
Especially, Source B has extensively been studied in the chemical survey project \citep{Jorgensen2016}. 

\newpage

\begin{wrapfigure}[51]{o}[0pt]{0.51\textwidth}
\begin{center}
\vspace*{-5pt}
\includegraphics[bb = 0 0 330 825, scale = 0.518]{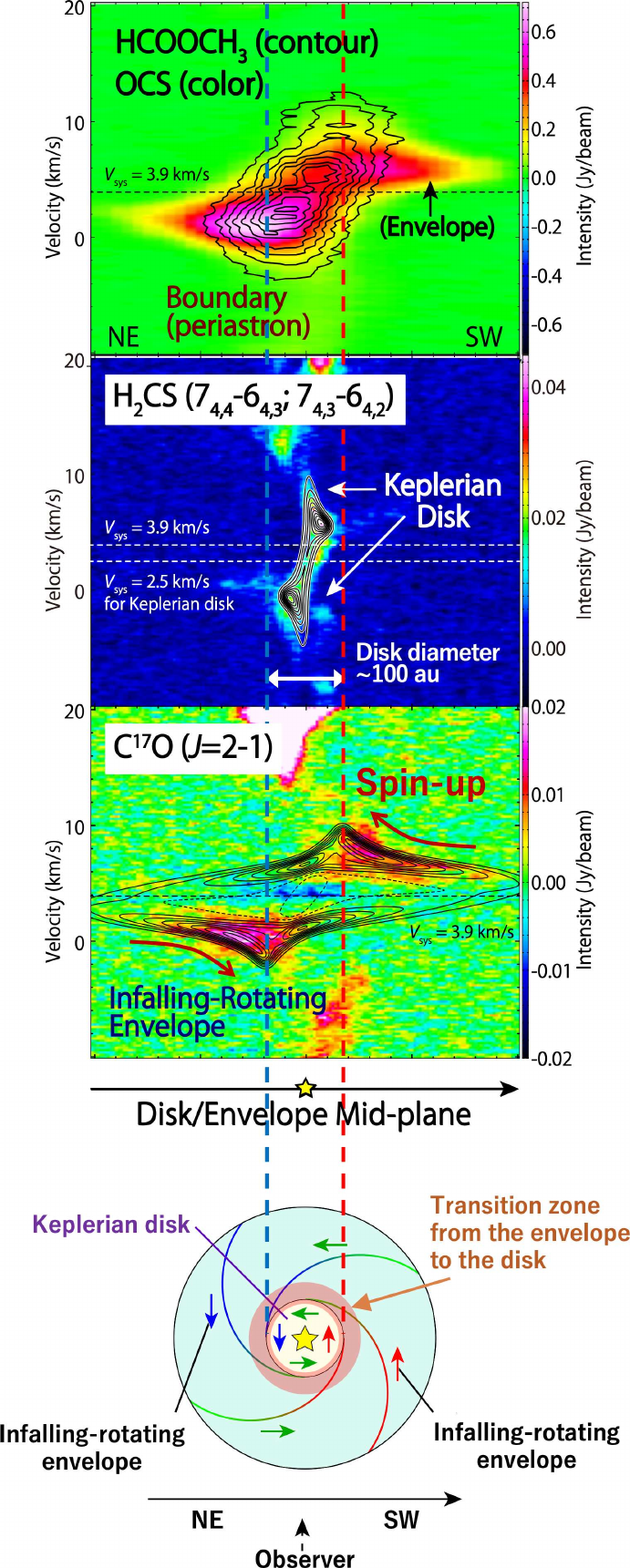}
\caption{%
Transition from the infalling-rotating envelope to the Keplerian disk observed in \iras\ Source A. 
Upper panels show the position-velocity diagrams of four molecular lines, 
where the position axis is taken along the mid-plane of the disk/envelope system. 
Bottom panel shows the schematic illustration of the disk/envelope system; 
the circummultiple structure surrounds Source A 
while the circumstellar disk is associated to the protostar A1 in Source A, 
although the multiplicity is omitted in this panel. 
The chemical composition of the gas drastically changes across the boundary between the envelope and the disk. 
Panels for the observational data 
are taken from \citet{Oya2016} and \citet{Oya2020} 
with modifications. 
}
\label{fig:16293}
\end{center}
\end{wrapfigure}

Studies for kinematic structures have been performed for Source A. 
Source A is reported to be a multiple source, 
consisting at least two protostars A1 and A2 \citep[e.g.][]{Chandler2005, Maureira2020}. 
The rotating gas in Source A has been reported \citep[e.g.][]{Pineda2012, Favre2014}. 
\citet{Oya2020} reported that 
the rotating structure is resolved into the circummultiple structure of Source A and the circumstellar disk of the protostar A1. 
They found that the chemical composition of the gas is different between these two physical structures. 
Figure~\ref{fig:16293} shows the position-velocity diagrams of molecular lines tracing the rotating motion in Source A. 
The \CO\ and OCS emission trace the circummultiple structure extending over 300 au, 
while the \TFA\ emission traces the circumstellar disk around the protostar A1. 
The kinematic structures of the circummultiple structure and the circumstellar disk 
can be approximated by an infalling-rotating motion and the Keplerian motion, respectively. 
The emission of \MN\ and \MF, which are characteristic to the famous hot corino activity of this source, 
were found to be locally enhanced in a ring-like structure with a radius of 50 au \citep{Oya2016}. 
It is interesting that 
this radius corresponds to the boundary between the circummultiple structure and the circumstellar disk.

A possible interpretation of the localized emission of COMs 
is their sublimation from dust grains by a local increase of the gas and dust temperature. 
The rotation temperature of \TFA\ was evaluated to be highest ($>300$ K) near the periastron of the circummultiple structure, 
and to be less than or comparable to 100 K inside and outside it \citep{Oya2016, Oya2020}, 
although this feature was not confirmed by \citet{Maureira2020}. 
This situation can be caused, for instance, 
by a weak accretion shock by the infalling gas and by radiation heating on a possible piled-up gas.  
If the dust temperature locally increases at the periastron of the circummultiple structure, 
it would help COMs to sublimate from dust grains there. 
That is, the chemical change is found to be attributed to the physical structure and the gas dynamics. 
This may be related to the origin of the famous hot corino activity in \iras\ Source A. 
Similar chemical differentiation was confirmed in 
\iras\ Source B \citep{Oya2018-16293B}, 
which was utilized to examine the kinematic structure in its almost face-on disk/envelope system.

\section{Overall View of Organic Molecules} \label{sec:overall}

\begin{figure}[!t]
\begin{center}
\includegraphics[bb = 0 0 900 350, scale = 0.4]{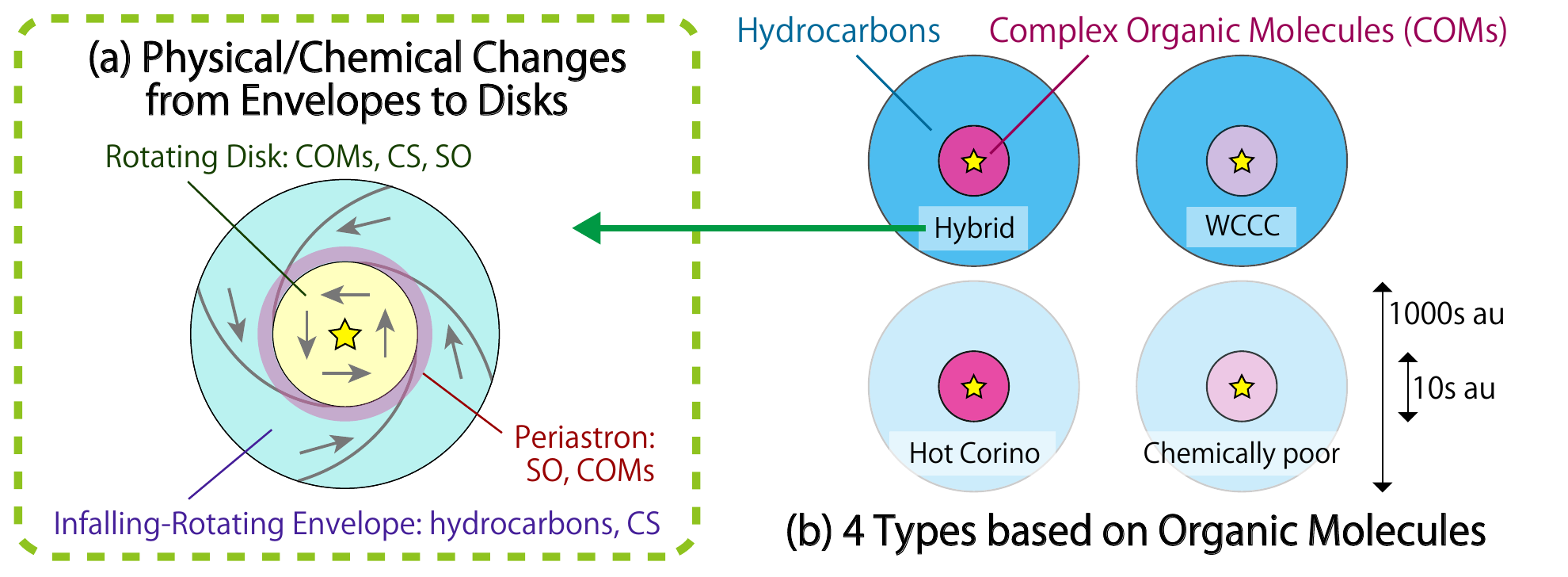}
\caption{%
(a) 
Schematic illustration of the relation between the physical structures and the chemical composition of the gas 
in a disk-forming region of a young low-mass protostellar source. 
See Section~\ref{sec:phys} for the detail. 
(b) 
Schematic illustration of the chemical differentiation and its diversity based on organic molecules in the gas-phase. 
See Sections~\ref{sec:chem-diff} and \ref{sec:chem-var} for the detail. 
}
\label{fig:summary}
\end{center}
\end{figure}

Our current understandings described in the previous sections can be summarized as Figure~\ref{fig:summary}. 
The study of the organic molecules in young low-mass protostellar sources had a large progress in this decade; 
the chemical diversity in protostellar cores found with single-dish telescopes can be spatially resolved by ALMA (Figure~\ref{fig:summary}b). 
WCCC occurs at a $10^3$ au scale, 
while hot corino chemistry occurs at a smaller scale in the vicinity of the protostar. 
The recent high spatial-resolution observations revealed that some sources show both the two chemical characteristics; 
the hybrid chemistry is likely the most standard case. 
Another extreme case is also found with neither the two kinds of characteristics even at a spatial resolution of 10s au \citep{Oya2019}. 
This is the overall trend from $10^{3-4}$ au scale down to a $10^{1-2}$ au scale.

It should be noted that the picture shown in Figure~\ref{fig:summary}(b) is merely a rough classification. 
The actual chemical composition need to be further examined 
not only by the presence/absence of the organic molecules in the gas phase 
but, of course, by to what amount of them exist in the gas- and ice-phase. 
Some specific study cases show 
that how extent the chemical characteristics can affect the apparent chemical composition. 
L483, which is identified to be a hybrid chemistry source with ALMA (Figure~\ref{fig:L483}), 
rather looks to be a carbon-chain rich source with the 45-m radio telescope at Nobeyama Radio Observatory \citep{Hirota2009}; 
the COM emissions on a 10s au scale is severely diluted with the beam size of $10^4$ au with the single-dish telescope. 
COM emissions have not firmly been detected toward \irass\ other than \MN\ \citep{Jorgensen2013, Oya2014, Okoda2021} 
even with the high spatial-resolution and sensitivity of ALMA observations. 
Nevertheless, \citet{Yang2022} revealed 
that various COMs are indeed produced in this source and trapped in the icy grain mantle 
based on their observation with Mid-Infrared Instrument (MIRI) on James Webb Space Telescope (JWST). 

%The change in the kinematic structure from the infalling-rotating envelope to the Keplerian disk is expected to be common in young protostellar sources. 
The chemical differentiation from an infalling-rotating envelope to a Keplerian disk found in \iras\ (Section~\ref{sec:phys}) 
has also been reported for other sources \citep[][and literatures therein]{Oya2022}. 
Moreover, 
the molecular species showing the chemical differentiation is different among sources depending on their chemical characteristics. 
The case for the hybrid source L483 is shown in Figure~\ref{fig:summary}(a). 
That is, the chemical diversity known in the parent cores is indeed delivered into the scale of the disk formation. 
The chemical change and its diversity can affect the material inheritance from the interstellar gas to protostellar disks, 
and possibly planets. 

The ALMA large project FAUST 
(Fifty AU STudy of the chemistry in the disk/envelope system of Solar-like protostars) is now being progressed 
to establish this unified picture of the physical/chemical view 
with more number of target sources and molecular species. 
As discussed so far, 
it should be emphasized that 
the molecular line emissions are deeply related to 
the physical condition {\it in the past} and {\it at present.}

\section{Transition Zone from Envelopes to Disks} \label{sec:transition}

In these years, 
the chemical view and the physical structures in protoplanetary disks, 
including their substructures, 
have extensively been examined, 
for instance in the ALMA large projects 
MAPS \citep[Molecules with ALMA at Planet-forming Scales;][]{Oberg2021} 
and DSHARP \citep[Disk Substructures at High Angular Resolution Project;][]{Andrews2018}. 
Meanwhile, our current understandings for younger evolutionary stages (Figure~\ref{fig:summary}) 
would be a too simplified picture. 
In reality, the situation can be much more complicated at a smaller scale. 
One of the current problem is 
what is occurring in the transition zone from accreting gas to rotaionally-supported disks.

\subsection{Variation among Molecular Species} \label{sec:transition_NO}
In \iras\ Source A (Section~\ref{sec:phys}), 
the chemical composition of the gas is different 
between the circumstellar disk of the protostar A1 
and the vicinity of the protostar A2. 
Its multiplicity may affect the chemical evolution 
from the accreting gas to protostellar disks. 
(The figure with preliminary results are dropped from this book.) 

Not only for multiple systems, 
a variety of radial chemical distributions have recently been recognized as a common occurrence. 
For instance, 
ring-like distributions in protoplanetary disks 
were revealed in the ALMA large project ALMA-DOT 
\citep[ALMA chemical survey of Disk-Outflow sources in Taurus;][]{Garufi2021}; 
some molecular species are thought to be destructed 
by molecules sublimated in the vicinity of the protostar 
and/or UV radiation. 
On the contrary, 
some molecular emissions are enhanced only near the protostar; 
observational works for Nitrogen-bearing organic molecules 
have extensively been reported in these years \citep{Nazari2021}. %\citep{vanGelder2020, Nazari2021}. 
Their distributions tend to be compact 
relative to the Oxygen-bearing organic molecules \citep{Okoda2021}. 
We need a higher-angular resolution to characterize 
the chemical change within disks structures. 
Moreover, 
asymmetric or localized distributions are also found near protostars. 
Gas dynamics is an important factor; 
as well as the accretion shocks at the periastron (Section~\ref{sec:phys}), 
streamers are now in the spot light \citep[e.g.][]{Harsono2023}. 

\subsection{Variation within One Species} \label{sec:transition_OCS}

%\begin{wrapfigure}[35]{o}[0pt]{0.6\textwidth}
\begin{figure}
\begin{center}
\includegraphics[bb = 0 0 520 520, scale = 0.6]{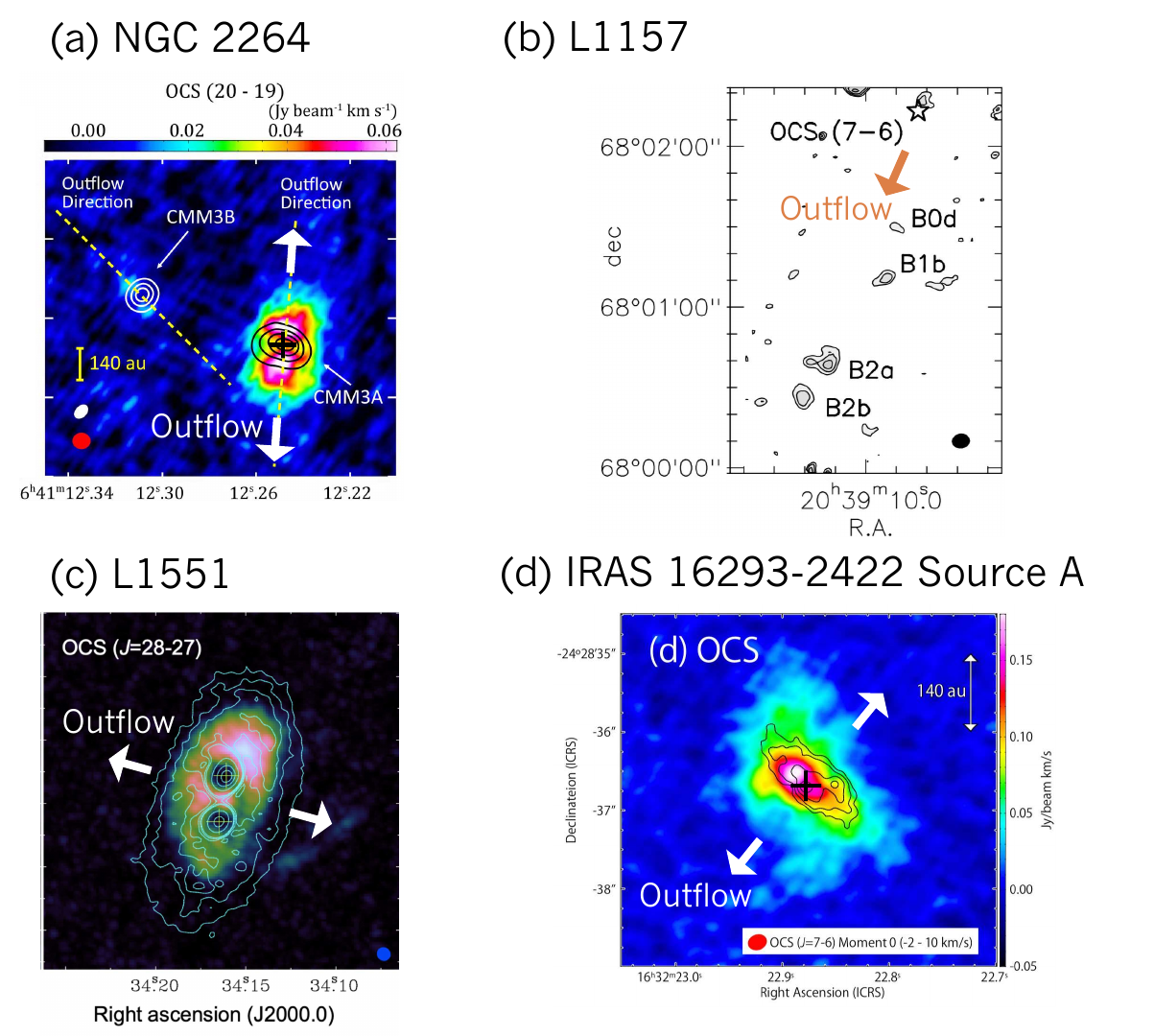}
\caption{%
Observations of the OCS emission in four different sources. 
(a) 
The OCS ($J=20-19$) emission 
is detected in the outflow of 
a Class 0 intermediate source NGC 2264 CMM3A. 
Taken from \citet{Shibayama2021}. 
(b) 
The OCS ($J=7-6$) emission 
is detected in the shocked regions by the outflow in L1157. 
Taken from \citet{Benedettini2007}. 
(c) 
The OCS ($J=28-27$) emission 
traces the circumbinary disk 
in L1551 IRS5. 
Taken from \citet{Takakuwa2020}. 
(d) 
The OCS ($J=7-6$) emission 
traces the circummultiple structure 
of a Class 0 protostellar source \iras\ Source A 
and is weakly detected in the outflow. 
Taken from \citet{Oya2021}. 
}
\label{fig:OCS}
\end{center}
\end{figure}

There is another complexity; 
the chemical composition does not always correspond to the physical structure one by one 
in the actual observational studies. 
OCS is one example, as shown in Figure~\ref{fig:OCS}. 
The OCS emission traces the outflow structure in NGC 2264 \citep{Shibayama2021}. 
Meanwhile, 
the OCS emission does not trace the outflow lobes in L1157, 
but is detected only in the shocked regions \citep{Benedettini2007}. 
In the L1551 and \iras\ Source A, 
the OCS emissions are mainly detected in their disk/envelope systems 
rather than the outflows 
\citep{Takakuwa2020, Oya2021}. 
SO is another example; 
the SO emission is known to be enhanced in shocked regions \citep[e.g. L1157 B1][]{Bachiller1997}, 
including the accretion shocks at the periastron. 
The SO emission traces the overall morphology of the outflow in \iras\ Source A \citep{Oya2021}, 
Meanwhile, 
it has recently been reported that 
the SO emissions can trace the disk/envelope systems 
\citep[B335, L483, \irass;][]{Imai2016, Oya2017, Okoda2018}.

It is well known that some specific molecules can work as tracers; 
for instance, 
SiO is thought to be a shock tracer, 
and categorization of some other species are suggested by \citet{Tychoniec2021} 
as well as shown in Figure~\ref{fig:summary}(a). 
However, 
some species can trace various structures depending on the characteristics of sources. 
We need careful consideration 
which molecular species trace which structure {\it in each source}. 
Sulfur-bearing species are tend to show this feature in disk-forming regions of protostellar sources, as described above. 
With this in mind, 
we are now conducting a survey with ALMA to reveal their distributions 
in young low-mass protostellar sources.

\section{Prospects for Next Decades} \label{sec:prospects}

The new questions discussed in the previous sections will be hot topics in the next decade. 

\subsection{How to Look into the Scale for the Planetary Formation?} \label{sec:prospects_planet}
\begin{figure}
\begin{center}
\includegraphics[bb = 0 0 430 150, scale = 0.82]{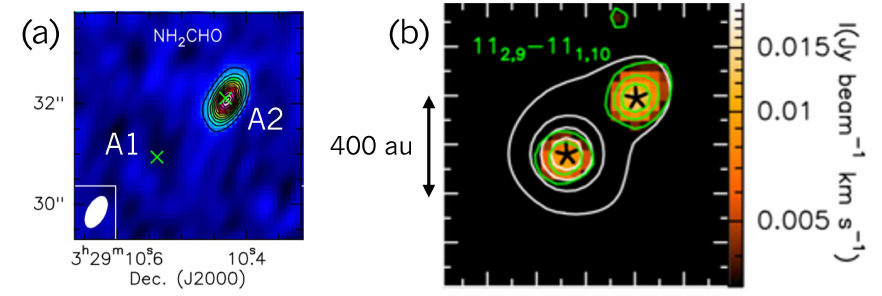}
\caption{%
Emission of COMs in IRAS 4A. 
(a) 
The NH$_2$CHO ($12_{0,12}-11_{0,11}$) emission 
is detected only in the vicinity of the protostar IRAS 4A2 with ALMA. 
Taken from \citet{LopezSepulcre2017} with modifications. 
(b) 
The \MN\ ($11_{2,9}-11_{1,10}$) emission 
is detected in the vicinity of both the two protostars with VLA. 
Taken from \citet{DeSimone2020}. 
}
\label{fig:IRAS4A1}
\end{center}
\end{figure}

As extensively discussed in this symposium, 
one important key to solve the complexity in the vicinity of the protostars 
is the ice chemistry (see also Section~\ref{sec:overall}). 
As well, 
the observations at the cm-wavelength will be essential.  
Development and enhancement of telescopes 
are expected to open a new avenue for this field; 
such as ALMA 2, Square Killometer Array (SKA), and next generation Very Large Array (ngVLA). 
We can achieve a spatial resolution even as high as a few au, 
however, 
the gas in such a closest vicinity of protostars is often opaque at mm/sub-mm wavelength. 
The case in IRAS 4A1 is shown as an example in Figure~\ref{fig:IRAS4A1}. 
The observation of the NH$_2$CHO emission seems 
as if the two protostars would have different chemical characteristics (Figure~\ref{fig:IRAS4A1}a). 
However, non-detection of this COM emission was explained 
by the attenuation due to the optically thick dust continuum emission at mm/sub-mm wavelength \citep{DeSimone2020}; 
in fact, multiple lines of \MN\ were indeed detected at cm-wavelength with VLA (Figure~\ref{fig:IRAS4A1}b).

Moreover, 
the chemical complexity will be further examined at cm-wavelength, 
because there are line emissions of larger and heavier molecular species 
in comparison to mm/sub-mm wavelength. 
We are expecting to directly look into the scale of the planetary formation 
with a wide variety of organic molecules.

\subsection{How to Cook the Huge Pile of Observational Data?} \label{sec:prospects_ML}

\begin{figure}
\begin{center}
\includegraphics[bb = 0 0 550 420, scale = 0.7]{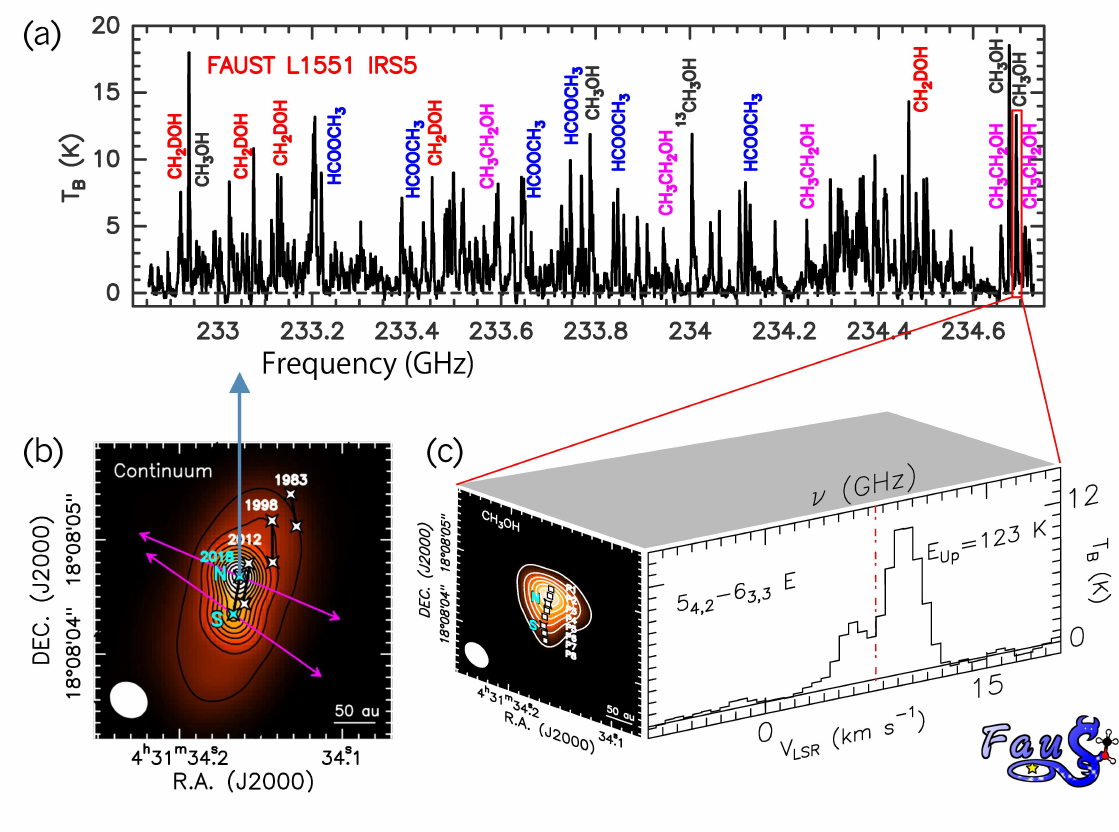}
\caption{%
Observation for L1551 IRS5 
as a part of the ALMA large project FAUST. 
Panels are taken from \citet{Bianchi2020} and modified. 
(a) 
Spectra taken toward the continuum peak IRS5 N. 
(b) 
Map of the 1.3 mm dust continuum emission. 
(c) 
Integrated intensity map of the \MN\ ($5_{4, 2}-6_{3, 3}$; E) emission 
and its spectral profile toward the continuum peak IRS5 N. 
}
\label{fig:L1551}
\end{center}
\end{figure}

Complex physical and chemical structures have recently been found in young low-mass protostellar sources, 
as described in the previous sections. 
To solve the problems in these complexities, 
one direction is to make full use of the huge amount of data. 
We have flood of observational data in these days; 
for instance, 
archival systems of observatories and the public data of large projects. 
Figure~\ref{fig:L1551} shows an example of numerous molecular lines 
detected at once with ALMA. 
We need to consider how to cook such a huge pile of data efficiently.

\begin{figure}
\begin{center}
\includegraphics[bb = 0 0 800 350, scale = 0.53]{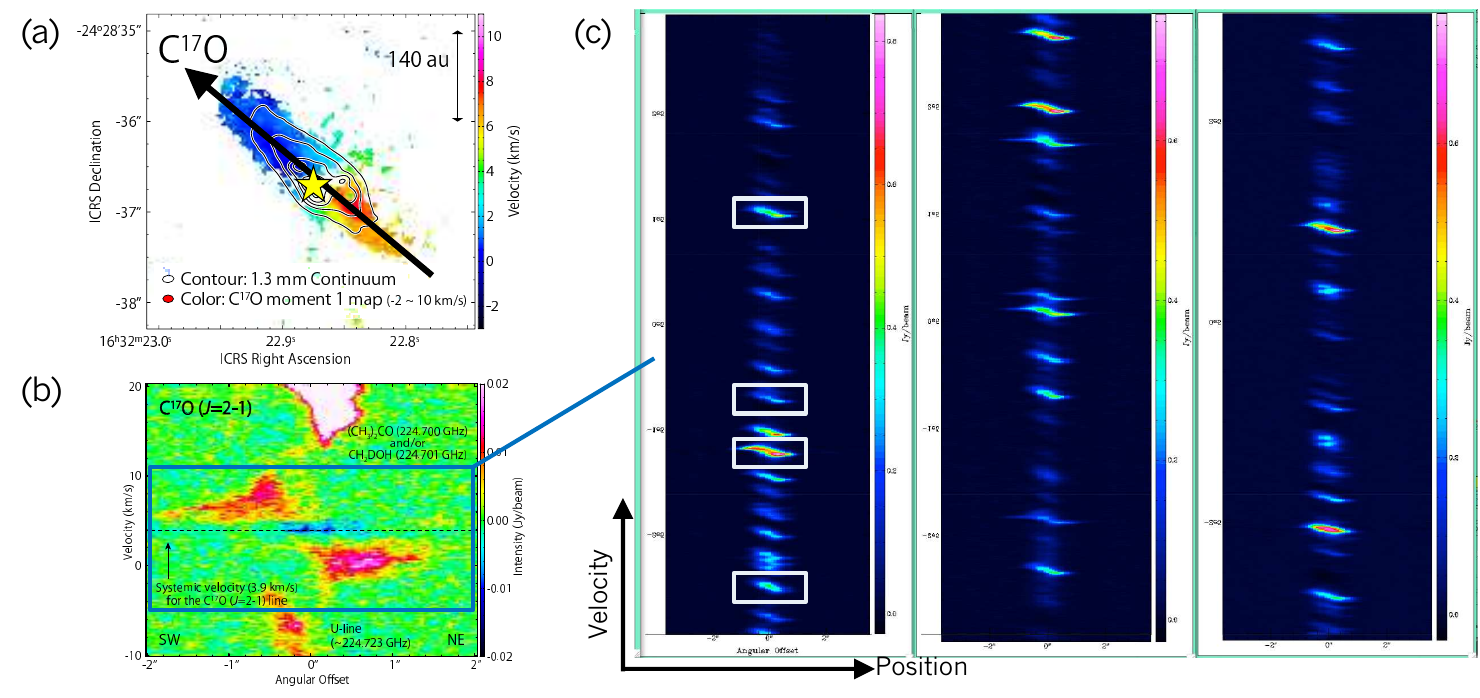}
\caption{%
Kinematic structure in \iras\ Source A. 
Panels (a) and (b) are taken from \citet{Oya2020} and modified. 
(a) 
Velocity map (moment 1 map) of the \CO\ emission. 
The velocity gradient is seen along the northeast-southwest direction 
due to the infall and rotation motion surrounding the multiple system. 
(b) 
Position-velocity diagram of the \CO\ emission. 
The position axis is taken along the mid-plane of the circummultiple structure, 
as represented by the arrow in panel (a). 
(c) 
Position-velocity diagram of three spectral windows 
prepared along the disk/envelope system. 
The observational data is taken from 
the PILS (The ALMA Protostellar Interferometric Line Survey) survey with ALMA \citep{Jorgensen2013}. 
}
\label{fig:16293-PV}
\end{center}
\end{figure}

As discussed in the previous sections, 
the studies of chemical compositions and the physical structures are inseparable. 
Then, a comprehensive examination in the three-dimensional cube data 
for each of these various molecular emissions are required 
to fully use the information obtained in observations. 
Figure~\ref{fig:16293-PV} shows the actual situation we have. 
In the position-velocity diagrams with a wide range of frequency (Figure~\ref{fig:16293-PV}c), 
we see similar tilted structures for many times along the mid-plane of the disk/envelope system (Figure~\ref{fig:16293-PV}a). 
Each of them corresponds to one molecular emission. 
Each molecular line traces the velocity gradient due to the gas motion, 
as seen in the \CO\ emission shown in Figure~\ref{fig:16293-PV}(b) as an example. 
We have many molecular emissions, 
and simultaneously, the information of their kinematic structures. 
It is, unfortunately, too hard to examine all this flood of information 
one by one manually. 

One possible solution to resolve this tough situation 
is applying machine and deep learnings. 
\citet{Oya2022} performed a supervised machine learning method; 
the observed cube-data of the molecular lines 
are classified into the infalling-rotating envelope and the Keplerian disk 
with the aid of the support vector machine. 
They demonstrated that 
the analysis in the kinematics can help us to interpret the observed chemical view. 

Conversely, 
the chemical diagnostics can be a powerful tool to analyze the gas dynamics. 
For instance, 
\citet{Oya2021} examined 
the re-distribution of the angular momentum of the gas during the disk formation 
among the accreting gas, the rotating disk, and the outflow. 
Rotating motion of outflows have been detected in several sources in these years 
\citep[e.g.][]{Hirota2017, Oya2018-L483, Zhang2018}.
Because a possible rotation motion of an outflow is expected to be most prominent  
near its launching point with the smallest radial size \citep{Oya2015}, 
it is essential to find a suitable molecular tracer 
to avoid a contamination from the disk/envelope component there. 
This approach is not limited for low-mass protostellar sources. 
There would be a chance to apply to other rotating structures, 
such as high-mass sources \citep{Olguin2023}
and even active galactic nuclei \citep{Aalto2020}. 
These rotating structures have much different spatial scales, 
but can have some analogies in physics. 
This may provide us with a clue to the structure formation with angular momentum.

\section{Summary} \label{sec:summary}

As discussed so far, 
we have some progresses to fill out the missing-link in the disk-formation study in these years. 
We see {\it the co-evolution of the chemical compositions and the physical structures}. 
As the molecular distributions are classified by comparing with simple physical models, 
the study in the chemical composition is inseparable from the information in the kinematics. 

The next step is looking into the planet-forming regions. 
As for the structure formation process, 
re-distribution of the angular momentum has an essential role 
to determine the initial condition for the planet formation. 
The chemical evolution along this gas dynamics will be 
an important clue to the material origin of planets. 
Meanwhile, 
the study in planetary systems are going to have extensive progresses, 
for example with Hayabusa-2 and JWST. 
We are still on the way to bridge 
our understandings from the star-formation study to the planetary science.

\end{document}